\documentstyle[aps,prd,eqsecnum]{revtex}
\begin{document}
\title{\bf ORDER PARAMETER EVOLUTION IN SCALAR QFT: RENORMALIZATION
GROUP RESUMMATION OF SECULAR TERMS} 

\author{\bf 
H. J. de Vega and  J. F. J. Salgado\footnote{e-mail:salgado@lpthe.jussieu.fr}}
\address
{ L.P.T.H.E.\footnote{Laboratoire Associ\'{e} au CNRS UA280.},
Universit\'es Pierre et Marie Curie (Paris VI) 
et Denis Diderot  (Paris VII), \\ Tour 16, 1\raisebox{1ex}{er} \'etage, 4 Place Jussieu, 
75252 Paris Cedex 05,  FRANCE}
\date{January 1997}
\maketitle

\begin{abstract}
The quantum evolution equations for the field expectation value are 
 analytically solved to cubic order in the field amplitude and to
 one-loop level in the $\lambda\phi^4$ model.
We adapt and use the renormalization group (RG) method  for such non-linear 
and non-local equations. The time dependence of the field expectation
 value is explicitly derived integrating the RG equations.
It is shown that the field amplitude for late times approaches  a
 finite limit as $ O(t^{-3/2}) $. This  limiting value is 
 expressed as a  function of the initial field amplitude. 
\end{abstract}

\section{Introduction and Motivations}

An important research activity develops presently on the non-equilibrium
dynamics of quantum field theory with multiple physical
 motivations\cite{nos1,dis,rusos,japos,otros,big}.
They arise in the  reheating of inflationary
universes, the eventual formation of
disordered chiral condensates \cite{dcc}, the  understanding the
hadronization stage of the 
quark-gluon plasma\cite{muller}  as well as trying to understand out of
equilibrium particle production in strong electromagnetic fields and in heavy
ion collisions \cite{mottola,habib}. 

A common feature in all such phenomena is the presence of
 quantum fields with large
amplitudes. That is, they imply the quantum field dynamics when 
the  energy density is {\bf high}. That is to say of order $ m^{-4} $
times a large number, where $ m $ is the typical mass scale in the theory.
 [For example,  $ \lambda^{-1}\; m^{-4} $,
 where $ \lambda $ is a coupling constant]. 
The usual S-matrix calculations apply in the opposite limit of low
energy density and since they only provide information on {\em in}
$\rightarrow$ {\em out} matrix elements,  are unsuitable for calculations of
expectation values. 

Our programme on non-equilibrium dynamics of quantum field theory, started in
 1992\cite{nos1}, is naturally poised to provide a framework to study these 
problems. Thorough and accurate numerical calculations involving
 large-$N$ and  Hartree  nonperturbative methods gave a clear
 picture of the physical processes involved both, qualitatively and
 quantitatively\cite{dis,nos3,big}. Very briefly, part of the energy
 initially stored in the zero mode is dissipated through quantum
 particle production. Since the initial field amplitude is typically
 of order $ 1/\sqrt{\lambda} $, this is a nonperturbative physical process.
Detailed analytical calculations yield precise results during the
 so-called preheating time. Namely, before non-linearities shut-off
 the process of particle production linked to parametric or spinodal
 instabilities \cite{big}. 

We consider here a $ \phi^4 $ scalar field theory in four space-time
dimensions (i.e. a standard inflationary model). We give the name of  order
parameter  to the quantum expectation value of the
field $\phi(t)$. We consider traslationally invariant quantum states with
constant (and finite) energy per unit volume. A nonlinear and nonlocal
equation of motion determines the time evolution of $ \phi(t) $
\cite{hu,nos1,dis}. To one-loop level and to cubic order in $ \phi(t) $, the
equation of motion can be written as\cite{dis}
\begin{eqnarray}
&& \ddot{\phi}(t)+m^2 {\phi}(t)+\frac{g}{6}
{\phi}^3(t)
 +  \frac{g^2}{4}{\phi}(t)\int^t_{t_0}dt'\; \phi(t')\; 
\dot{\phi}(t') \int \frac{d^3k}{(2\pi)^3}
\frac{\cos[2\omega_k(t-t')]}{2\, \omega^3_k} = 0 \; .
\label{eqmov}
\end{eqnarray}
where $ g $ and $m^2$ stand for the renormalized coupling constant
and mass and $ \omega_k=\sqrt{m^2+k^2} $. Eq.(\ref{eqmov}) is to be
supplemented by the initial data $ \phi(t_0) $ and  $ {\dot\phi}(t_0) $.

The purpose of this paper is to solve analytically the equation of
motion (\ref {eqmov}) within the renormalization group approach.
We obtain in this way analytic
expressions for the order parameter   at {\bf all times}.

Our final results can be summarized as follows. The order parameter
takes the form
$$
\phi(t) = \sqrt{{6m^2}\over g}\; \eta(mt)
$$
where
$$
\eta(t) = R(t) \; {\rm cn}(t\sqrt{1+R(t)^2},k(t)) \; ,
$$
where cn stands for the Jacobi elliptic cosinus and $ R(t) $ and $ k(t)
\equiv \frac{R(t)}{\sqrt{2(1+R(t)^2)}} $ vary slowly with time. 
[Slow compared with the variation of the elliptic function whose
period is of order one]. 

The time dependence of $ R(t) $ is determined by the renormalization group
equations. We give in eq.(\ref{final}) the explicit time dependence. 

We find for late times,
$$
R(t)-R_\infty
=\frac{\lambda\sqrt\pi}{32}\frac{R_\infty^3}{1-3R_\infty^2}
\frac{\Omega\cos\Omega t \; \cos(2t+\pi/4)+\sin\Omega t
\; \sin(2t+\pi/4)}{(\Omega+2)^2(\Omega-2)\; t^{3/2}} +O(t^{-2})
$$

That is, the field amplitude for late times approaches  a
 finite limit as $ O(t^{-3/2}) $. This  limiting value can be
 expressed as a  function of the initial amplitude $ R_0 $ by
$$
F(R_\infty)-F(R_0)=-\frac{\pi^5\lambda}{\omega_1^2}\left\lbrack
\frac{\omega_1^2+2}{\omega_1\sqrt{\omega_1^2-1}}
{\rm Arg}\cosh\omega_1-2\right\rbrack \; ,
$$
where the trascendental function $ F(R) $ is defined by
eqs.(\ref{funF}) and (\ref{Fexpli}).

For small initial amplitude this reduces to 
$$
R_\infty\simeq R_0\biggl(1-\lambda\frac{R_0^2}{8\pi}\biggr) +
{\cal O}(\lambda R_0^4)
$$

In summary the order parameter oscillates as the classical cnoidal
solution with slowly time dependent amplitude and frequency. The
amplitude decreases with time due to the energy dissipated in produced
particles. These results are in agreement with the numerical
calculations \cite{nos1,dis}.

\section{A pedagogical example}

The renormalization group method (RG) \cite{gold_90,gold_95,kunihiro}
 improves the knowledge
of the asymptotic behaviour of solutions of non-linear differential equations.
The long time power law behaviour  for non-linear
equations can be found by RG methods \cite{gold_90,bricmont}.
The RG not only  resums the secular terms that  
arise in an usual (na\"{\i}ve) perturbative expansion but also
yields an uniform expansion of the solution.
More, it is possible to choose between the different resumed approximations.
The different choices are, of course, all equivalent at a given order.
We will show that the choice of the final form is given from the choice of
the counterterms that are needed in the method.
Moreover, after one has choosen the final form of what can be called the 
envelope of the solution \cite{kunihiro}, one still 
has a freedom in the choice of the counterterms  and can introduce an
arbitrary function expressing  a kind of gauge freedom.
This gauge function may be used to give a simple expression
to  the slowly varying pieces. Of course, the final solution is 
independent of this gauge function.

In order to introduce these ideas we consider the exactly solvable example of 
the anharmonic oscillator with given initial  conditions:
\begin{eqnarray}
\ddot y+y+\epsilon y^3&=&0 \cr
y(t_0)=R_0,\ \ \dot y(t_0)&=&0 \nonumber
\end{eqnarray}
Where the exact solution is given in terms of elliptic functions \cite{gr}
\begin{equation}\label{solex}
y(t)=R_0 \; {\rm cn}\left((t-t_0)\sqrt{1+\epsilon R_0^2},k(\sqrt\epsilon
R_0)\right), \qquad 
k(x)=\frac x {\sqrt{2(1+x^2)}}
\end{equation}
Starting from the zeroth order solution (for $\epsilon=0$)
$$
y_0(t)=R_0\cos(t-t_0)
$$
one can easily get the first order solution 
by linear perturbation (or by expanding eq.(\ref{solex}) to order $
\epsilon $) and see that it contains a secular term (in $t-t_0$):
\begin{eqnarray}
\label{app_sol}
y(t)=& &R_0 \; \cos(t+\Theta_0)-\epsilon\; \frac 3 8
(t-t_0)\; R_0^3\; \sin(t+\Theta_0)\cr 
&+&\epsilon\;\frac{R_0^3}{32}\biggl[\cos 3(t+\Theta_0)-\cos(t+\Theta_0)\biggr]
+{\cal O}(\epsilon^2)\cr
\Theta(t_0)=&-&t_0
\end{eqnarray}
Now following the method described in ref.\cite{gold_95} we introduce the 
renormalization of $R_0$ and $\Theta_0$:
\begin{eqnarray}
R_0(t_0)&=&R(\tau)Z_1(t_0,\tau)\cr\cr
\Theta_0(t_0)&=&\Theta(\tau)+Z_2(t_0,\tau)\nonumber
\end{eqnarray}
Where $\tau$ is a constant which we shall choose later, we expand $Z_1$ and
$Z_2$ in power of $\epsilon$, after we we have noticed that $Z_1$ is a 
multiplicative renormalization and that $Z_2$ is an additive one:
\begin{eqnarray}
Z_1=&1+&\epsilon a_1 + {\cal O}(\epsilon^2)\cr\cr
Z_2=& &\epsilon b_1 + {\cal O}(\epsilon^2)\nonumber
\end{eqnarray}
Inserting these expansions in (\ref{app_sol}) we get:
\begin{eqnarray}
\label{app_ct}
y(t)=& &R\cos(t+\Theta)+\epsilon \; R  \;a_1\cos(t+\Theta)
-\epsilon \; R \; b_1\sin(t+\Theta)\cr
&-&\epsilon \;\frac 3 8 (t-\tau) \;R^3 \;\sin(t+\Theta)
-\epsilon \;\frac 3 8 (\tau-t_0) \;R^3 \;\sin(t+\Theta)\cr
&+&\epsilon \;\frac{R^3}{32}\biggl[\cos 3(t+\Theta)-\cos(t+\Theta)\biggr]
+{\cal O}(\epsilon^2)
\end{eqnarray}
Now we want to remove the term that grows with $ t_0 $, because for a fixed
(eventually large) $ t $, this is a secular term.
Although there are no ultraviolet neither infrared divergences, which 
are typical of quantum field theory, subtracting
such terms is the analogue of eliminating the divergent terms in a
quantum field theory calculation ($t_0 \leftrightarrow \ln \Lambda$),
where $ \Lambda$ is the ultraviolet cutoff. We call 
minimal subtraction scheme the one with the following
counterterms
\begin{eqnarray}
a_1^{MS}&=&0\cr \cr
b_1^{MS}&=&\frac 3 8 (t_0-\tau)\nonumber
\end{eqnarray}
Requiring that $y(t)$ {\it do not} depend on $\tau$ (as bare correlation 
functions do not depend on the renormalization scale in quantum field theory),
 we obtain:
$$
0=\frac{dy}{d\tau}(t)=\dot R(\tau)\cos(t+\Theta(\tau))
+\epsilon R\biggl[\frac 3 8 R(\tau)^2-\dot\Theta\biggr]\sin
\biggl(t+\Theta(\tau)\biggr)
+{\cal O}(\epsilon^2)
$$
Using the fact that for a fixed $\tau$, $\cos(t+\Theta(\tau))$ and
$\sin(t+\Theta(\tau))$ are independent functions; one can see that 
they are both
solution of the zeroth order equation. This yields   the
 renormalization group equations, (which can be obtained using different 
but equivalent means \cite{kunihiro2,ignigo}):
\begin{eqnarray}
\dot R&=&0+{\cal O}(\epsilon^2)\cr
\dot\Theta&=&\epsilon\; \frac 3 8 R^2+{\cal O}(\epsilon^2)
\end{eqnarray}
Now we consider that $R$ and $\Theta$ are solutions of the RG equations, hence
$y(t)$ is $\tau$-independent and we can choose $\tau$ arbitrarily. 
We choose $\tau=t$ and we recover the same result as expanding the exact
elliptic solution (see ref. \cite{gr,pbm}). (This approximation can be
derived by other methods, see ref. \cite{jordan}):
\begin{eqnarray}
\label{app_res1}
y(t)&=&R_0\cos\omega(t-t_0) 
+\epsilon\;\frac{R_0^3}{32}\biggl[\cos 3\omega(t-t_0)-\cos\omega(t-t_0)\biggr]
+{\cal O}(\epsilon^2)\cr
\omega&=&1+\epsilon \;\frac 3 8 R_0^2+{\cal O}(\epsilon^2)
\end{eqnarray}
Here the secular term have been resumed [compare with eq.(\ref{app_sol})].

We could have chosen other counter-terms, for instance:
\begin{eqnarray}
a_1&=&a_1^{MS}-\epsilon\; \frac{R^2}{32}\biggl[-1+\cos 2(\tau+\Theta)-g(\tau+\Theta)\sin(\tau+\Theta)\biggr] \cr
b_1&=&b_1^{MS}-\epsilon\;\frac{R^2}{32}\biggl[\sin 2(\tau+\Theta)
-g(\tau+\Theta)\cos(\tau+\Theta)\biggr]\nonumber
\end{eqnarray}
where $g(u)$ is an arbitrary function. One has to notice that the terms in
$\sin 2u$ and $\cos 2u$ are introduced to
 remove the term in $\cos 3(t+\Theta_0)$
in the first order solution (\ref{app_sol}).\\
When we insert these values in (\ref{app_ct}) we get:
\begin{eqnarray}
y(t)=&&R\; \cos(t+\Theta)+\epsilon\;\frac 3 8\;(\tau-t)R^3\sin(t+\Theta)\cr
&+&\epsilon\;\frac{R^3}{32}\biggl[\cos 3(t+\Theta)
-\cos(t+\Theta)\cos 2(\tau+\Theta)+\sin(t+\Theta)\sin 2(\tau+\Theta)+\cr
&&g(\tau+\Theta)(\sin(\tau+\Theta)\cos(t+\Theta)-
\cos(\tau+\Theta)\sin(t+\Theta)
\biggr]+{\cal O}(\epsilon^2)\nonumber
\end{eqnarray}
The RG equations take now the form:
\begin{eqnarray}
\dot R&=&-\epsilon\;\frac{R^3}{32}\Biggl\{2\sin 2(\tau+\Theta)+
\frac d{d\tau}\biggl[g(
\tau+\Theta)\sin(\tau+\Theta)\biggr]\Biggr\}+{\cal O}(\epsilon^2)\cr
\dot\Theta&=&\epsilon\;\frac 3 8 R^2+\epsilon\frac{R^2}{32}
   \Biggl\{2\cos 2(\tau+\Theta)-\frac d{d\tau}\biggl[
g(\tau+\Theta)\cos(\tau+\Theta)\biggr]\Biggr\}
+{\cal O}(\epsilon^2)\nonumber
\end{eqnarray}
If $R(t)$ and $\Theta(t)$ are  solutions of these
 last RG equations, $y(t)$ 
is  $\tau$-independent, but might depend on $g(u)$. In fact this is 
not the case, if one puts $\tau=t$ and expands the final solution to see
how it depends on $g(u)$ one notices that at this order all the terms 
containing $g(u)$ cancel. In other words, 
the final solution {\bf do not}
depend on the choice of this function $g(u)$.
The arbitrariness of the function $g(u)$ is like a gauge freedom in the present
context. We can use this gauge freedom to simplify the RG equations.\\
In the present case a good choice is:
$$
g(u)=-2\sin u
$$
With this value of the gauge function the RG equations become:
\begin{eqnarray}
\dot R&=&0 +{\cal O}(\epsilon^2)\nonumber\\
\dot \Theta&=&\epsilon\; \frac 3 8 R^2+\epsilon\; \frac{R^2}8\cos
2(\tau+\Theta) 
+{\cal O}(\epsilon^2)\nonumber
\end{eqnarray}
It is clear that if we again choose  $\tau=t$, all terms will 
vanish except the 
first one:
$$
y(t)=R(t)\cos(t+\Theta(t))+{\cal O}(\epsilon^2)
$$
We finally get:
\begin{eqnarray}
y(t)&=&R_0\cos\Biggl[\omega(t-t_0)+\epsilon\frac{R^2}{16}\sin
2\omega(t-t_0)\Biggr]
+{\cal O}(\epsilon^2)\cr\cr
\omega&=&1+\epsilon\frac 3 8 R_0^2+{\cal O}(\epsilon^2)\nonumber
\end{eqnarray}
Expanding this last result in powers of $\epsilon$
we recover the previous equation (\ref{app_res1}).

\section{Out of equilibrium quantum field theory}

 Let us  consider the   self-coupled $\phi^4$ scalar
field theory with Lagrangian density
\begin{equation}
{\cal{L}} = \frac{1}{2}\partial_{\mu}\Phi\partial^{\mu}\Phi-
\frac{1}{2}m^2\Phi^2-\frac{g}{4!}\Phi^4 \label{Lagrangian}
\end{equation}
The time evolution of the  order parameter defined as the
 quantum expectation value,
\begin{equation}
\phi(t) = \langle \Phi(x,t) \rangle
\label{ordpar}
\end{equation}
is governed by non-linear and non-local equations.
We consider  translational invariant quantum states.\\
As shown in ref.\cite{dis}, to order $ g $ (one-loop level)
and keeping the first
non-linear contribution in the field amplitude $ \phi(t) $, the equation
of motion for $ \phi(t) $ takes the form:

\begin{equation}
 \ddot{\phi}(t)+m^2 {\phi}(t)+\frac{g}{6}
{\phi}^3(t)
 +  \frac{g^2}{4}{\phi}(t)\int^t_{t_0}dt'\; \phi(t')\; 
\dot{\phi}(t') \int \frac{d^3k}{(2\pi)^3}
\frac{\cos[2\omega_k(t-t')]}{2\, \omega^3_k} = 0 \; .
\label{eqII}
\end{equation}
Here  $ g $ and $m^2$ stand for the renormalized coupling constant
and mass, respectively. We have chosen a finite time $t_0$  as the beginning
of the self-coupling interaction as in ref. \cite{dis}.

It must be noticed that  the last term in   eq. (\ref{eqII}),
(the `dissipative' contribution), has a non-Markovian (i.e. memory-retaining)
kernel. Secondly, the equation is {\it non-linear} in the field
amplitude  $\phi(t)$. Moreover, as shown in \cite{dis}, a   $\Gamma
\dot{\phi}$ term in the equation of motion {\bf cannot} reproduce the dynamics.

Now, we introduce  dimensionless variables and replace $t$ by $t/m$
$$
\eta(t)    =  \sqrt{\frac{g}{6 m^2}} \;\phi({t \over m})  
\; \; ; \; \;  
q        =  \frac{k}{m} \; \; ; \; \; 
\lambda = \frac{3}{8\pi}\, g  \; 
$$
and eq.(\ref{eqII}) becomes
\begin{equation} \label{fieldmotion}
\ddot{\eta}+ \eta+ \eta^3+ \lambda\, h[\eta] = 0\; , 
\end{equation}
where
\begin{eqnarray}\label{nucleo}
 h[\eta(t)]&=&\eta(t) \int_{t_0}^t  \dot\eta(t')  \eta(t) K(t'-t) dt' \cr\cr
 K(t)&=&\int_0^{\infty} \frac{ k^2 dk}{ 2\pi} \frac{\cos
2{\omega}t}{\omega^3} \quad {\rm and} \quad \omega^2\equiv 1+k^2
\end{eqnarray}
Since $t_0$ is the beginning of the out of equilibrium evolution
the  initial conditions for $t = t_0$ are given as in ref. \cite{dis}
$$
\eta(t_0) = R_0  \quad {\rm and} \quad  \dot\eta(t_0) =0
$$
In order to understand the late time behaviour of the order parameter, we
shall solve this non-linear equation (\ref{fieldmotion}).
Using the renormalization groups methods \cite{gold_90,gold_95,kunihiro}
we will derive this behaviour in term of the initial conditions, up to 
the first order in $\lambda$ and third order in the amplitude.

\section{The na\"{\i}ve  perturbative expansion}

We start by seeking a standard perturbative solution in powers of  $\lambda$,
\begin{eqnarray}\label{pertu}
 \eta(t) &=& \eta_0(t)+\lambda\, \eta_1(t) +{\cal O}(\lambda^2) \;
 ,\cr \cr
 {\rm with}\quad \eta_0(t_0) &=& R_0  \quad {\rm and} \quad
 \dot\eta_0(t_0) = 0 \; .
\end{eqnarray}
We have then to solve to zeroth order:
\begin{eqnarray} \label{fimo}
 \ddot{\eta_0}+ \eta_0+ \eta_0^3 &=& 0  \; , \cr \cr
 \eta_0(t_0) &=& R_0   \quad {\rm and} \quad
 \dot\eta_0(t_0) =0\; .
\end{eqnarray}
The first order correction obeys,
\begin{eqnarray}
 \ddot{\eta_1}+ \eta_1+3 \eta_0^2\eta_1+h[\eta_0] &=& 0  \; ,
\label{eqneta1}\cr \cr
 \eta_1(t_0) &=& 0  \quad {\rm and} \quad
 \dot\eta_1(t_0) = 0\; .
\end{eqnarray}
We now introduce the function:
$$
f(t,R)=R\ {\rm cn}(t\sqrt{1+R^2},k(R)) \; ,
$$
with
\begin{equation}\label{kdeR}
k(R)=\frac{R}{\sqrt{2(1+R^2)}} \; .
\end{equation}
It is straightforward to show \cite{gr} that $f(t,R)$ obeys
eq.(\ref{fimo}) for $t_0=0$. 
\begin{eqnarray}
 \ddot{f}+f+f^3 &=& 0  \; ,\cr \cr
 f(0) &=& R \quad , \quad  \dot{f}(0) = 0 \; .\nonumber
\end{eqnarray}
So we have to zeroth-order the (purely classical) solution:
\begin{equation}\label{fimoC}
\eta_0(t)=f(t-t_0,R_0) \; .
\end{equation}
In order to find the one-loop quantum correction $\eta_1$, 
we have to solve the linear differential equation (\ref{eqneta1}). We define:
$$
 f_1(t,R)\equiv \frac{\partial f}{\partial t}(t,R)  \quad , \quad 
 f_2(t,R)\equiv\frac{\partial f}{\partial R}(t,R) \; .
$$
As $f(t)$ is a solution of (\ref{fimo})
and $f_1(t)$ and $f_2(t)$ are both derivatives of $f(t)$,
 $f_1(t)$ and $f_2(t)$ are solutions of
the homogeneous part of (\ref{eqneta1}).
Using the Green's function method, we can write the solution of
eq.(\ref{eqneta1}) as
\begin{equation}\label{fimo1}
\eta_1(t)=\frac{1}{W_0}\left\lbrack f_2(t-t_0)\int_{t_0}^t
f_1(t'-t_0)h(t')dt'- f_1(t-t_0)\int_{t_0}^t
f_2(t'-t_0)h(t')dt'\right\rbrack  \; . 
\end{equation}
Where $ h(t) $ stands for $ h[\eta_0(t)] $ and where $ W_0 $ is the Wronskian
of $ f_1(t,R) $ and $ f_2(t,R) $: 
$$
W_0\equiv W[f_1(t,R), f_2(t,R)]=-R_0(1+R_0^2) \; . 
$$
We now have the first two terms of the  perturbative expansion of
$\eta(t) $ in powers of $ \lambda $ . Unfortunately, they
contain {\bf secular} terms that make the expansion (\ref{pertu}) useless
for  $t\geq\lambda^{-1}$.
To see this, one can Fourier expand the periodic functions $f(t)$ and $f_1(t)$
(see \cite{gr}) as: 

\begin{eqnarray}
f(t,R) &=& \sum_{n=0}^{\infty} f_n\, \cos (2n-1) \Omega t \cr \cr
f_1(t,R) &=& -\Omega \sum_{n=0}^{\infty} (2n-1)\, f_n\, \sin (2n-1)
\Omega t \cr \cr
\Omega & \equiv & \frac{\pi}{K(k)} \sqrt{1+R^2} ,\qquad
f_n \equiv  \frac{2 \pi R}{k\ K(k)} \frac{q^{n-1/2}}{1+q^{2n-1}} \; ,\qquad
q  \equiv e^{-\pi K'(k)/K(k)}  \; . \label{constant}
\end{eqnarray}

One can notice that the function  $f_2(t)$ is not periodic in $t$.
However, the function  $f_3(t)$
defined as
$$
f_3(t) \equiv f_2(t) - t\; \frac{\Omega'}{\Omega}f_1(t) \; , 
$$
is a  periodic function of  $t$ with the  Fourier expansion
$$
f_3(t)=\sum_{n=0}^{\infty} {f'}_n \cos (2n-1) \Omega t \; .
$$
Here $ ' $ stands for $\frac{\partial }{\partial R}$.

This shows that $f_2(t)$ contains a secular term, that will contribute
through  eq.(\ref{fimo1}), to the secular terms in  $\eta_1(t)$.
The integrals of  $ f_1(t-t')h(t') $ and $ f_2(t-t')h(t') $ in
eq.(\ref{fimo1}) yield secular terms. The second expression produces a
second order secular term (growing as  $ t^2 $).

\section{Renormalization group equations}

In order to analyze the late time behaviour of the parameter,
we want to recast $\eta(t)$ in the form:
$$
\eta(t)=f(t + \Theta(t), R(t)) \; ,
$$
where the $t$ dependence of $ \Theta $ and $ R $ is slow. That is, 
${\dot  \Theta} = {\cal O}(\lambda)\; , \; {\dot R} = {\cal O}(\lambda) $. 
Since $f(t)$ is an amplitude $R$ multiplied by an oscillatory function
(the elliptic cosine), $R(t)$ will be the envelope of the solution.

In order to do that, we write the parameters $R_0$ and  $ \Theta_0(t_0) \equiv
-t_0$ in terms of new parameters $ \Theta $ and $ R $ through 
(finite) renormalizations:
\begin{eqnarray}
 R_0(t_0) &=& R(\tau)\ Z_1(t_0,\tau) \cr\cr
 \Theta_0(t_0) &=& \Theta(\tau)+Z_2(t_0,\tau) \nonumber
\end{eqnarray}
Here $\tau$ is a (new) constant parameter and
$Z_1$ and $Z_2$ have the following expansion in power of $\lambda$,
which ensure that both $R$ and $\Theta$ are of order ${\cal O}(\lambda)$: 
\begin{eqnarray}
  Z_1=1+\lambda\, a_1(t_0,\tau) +{\cal O}(\lambda^2) \cr\cr
  Z_2=0+\lambda\, b_1(t_0,\tau) +{\cal O}(\lambda^2)\nonumber
\end{eqnarray}

The requirement that $\eta(t) =  \eta_0(t)+\lambda\, \eta_1(t) +{\cal O}
(\lambda^2)$
takes the form $f(t + \Theta(\tau), R(\tau)) $ up to ${\cal O}(\lambda^2)$ fixes 
the (finite) counterterms $a_1$ and $b_1$ up to an arbitrary `gauge' function
$g(\tau)$. We find using eqs.(\ref{fimoC}) and (\ref{fimo1}),

\begin{eqnarray}
 a_1(t_0,\tau) = & & \frac{g(\tau)}{W} f_1(\tau+\Theta) -
\frac{1}{W} \int_{t_0}^{\tau}f_1(t'+\Theta) h(t') dt'  \quad , \cr \cr
 b_1(t_0,\tau) = & & -\frac{g(\tau)}{W}f_3(\tau+\Theta) -a_1(t_0,\tau)
\, (\tau-t_0) \frac{\Omega'}{\Omega} \cr  \cr
& &- \frac{1}{W} \left\lbrack 
\frac{\Omega'}{\Omega}(\tau-t_0)\int_{t_0}^{\tau}f_1(t'+\Theta) h(t') dt'-
\int_{t_0}^{\tau}f_2(t'+\Theta) h(t') dt'\right\rbrack \quad .\nonumber
\end{eqnarray}

Where we have written $\Theta$ for $\Theta(\tau)$, $R$ for $R(\tau)$ and 
$f_i(t)$ for $f_i(t,R(\tau))$. We  find, 
\begin{eqnarray}
\eta(t)=& & f(t+\Theta,R) - \lambda \; \frac{g(\tau)}{W}\left\lbrack 
f_1(t+\Theta)f_3(\tau+\Theta)-f_3(t+\Theta)f_1(\tau+\Theta)\right\rbrack \cr\cr
& & +\lambda  \; \frac{g(\tau)}{W} f_1(\tau+\Theta)f_1(t+\Theta)
\frac{\Omega'}{\Omega}(t-\tau)\cr\cr
& & +\frac{\lambda}{W}\left\{ \biggl[ f_3(t+\Theta)+(t-t_0)
\frac{\Omega'}{\Omega}f_1(t+\Theta)\biggr]
\int_{\tau}^{t}f_1(t'+\Theta) h(t') dt'   \right.\cr\cr
& & \left. -f_1(t+\Theta)\int_{\tau}^{t}f_2(t'+\Theta) h(t') dt' \right\} \;
 + {\cal O}(\lambda^2) \label{eta2} \; .
\end{eqnarray}

Since $\eta(t)$ is clearly $\tau$ independent:
\begin{equation}\label{tauin}
 \frac{d \eta}{d \tau}(t)=0 \quad ,\qquad \forall t \quad .
\end{equation}

Imposing eq.(\ref{tauin}) to eq.(\ref{eta2}) yields 
 the renormalization group equations to order $\lambda$.

\begin{eqnarray} \label{RG}
\dot{R}(t) &=& \ \ \frac{\lambda}{W} \left\{h(t)f_1(t+\Theta(t))- 
\frac{d}{dt}\biggl[g(t)f_1(t+\Theta(t))\biggr]\right\} \cr  \cr
\dot{\Theta}(t) &=&  -\frac{\lambda}{W}  \left\{ h(t)f_2(t+\Theta(t))- 
\frac{d}{dt}\biggl[ g(t)f_2(t+\Theta(t)) \biggr]\right\} \quad .
\end{eqnarray}
In the calculations, we used the fact that $\dot{\Theta}$ and 
$\dot{R}$ are of order ${\cal O}(\lambda)$
and the linear independence of $f_1$ and $f_2$.

Since $\tau$ is arbitrary and  $\eta(t)$ is  $\tau$ independent we can set
 $\tau$ equal to $t$ in eq.(\ref{eta2}). We thus find,
$$
\eta(t)=f(t+\Theta(t),R(t))+{\cal O}(\lambda^2) \; .
$$
We have thus succeeded to resum the secular terms.

\section{Fixing the `gauge' function }

We introduced an arbitrary function  $ g(\tau) $ in the counter terms $a_1$ and
$b_1$. This function  appears in the renormalization group equations.
The solution $ \eta(t) $ must be independent of the choice of  $ g(\tau) $.
This is easy to prove to this order.
One can choose the freedom in $g$ to give to  $\Theta$
or $R$ a specific form. It is clear that $\dot\Theta$ is related to
the finite  mass  renormalization and so has to be bounded for large $t$. 

We now study $\dot{\Theta}$ from eq.(\ref{RG}). We have to compute 
$$
h(\tau)= f(\tau+\Theta)\int_0^\tau dt'\
f\dot{f}(t'+\Theta)\int\ d\mu(\omega)\ \cos 2 \omega (t'-\tau) \; .
$$
Where the integration measure for $\omega$ is
$$
d\mu(\omega)=\frac{k^2}{2\pi \omega^3}\,dk=\frac{\sqrt{\omega^2-1}}{2
\pi \omega^2}\,d\omega 
$$
We introduce
$$
I(t)=\int_0^t dt'\ f\dot{f}(t'+\Theta)\int\ d\mu(\omega)\ \cos 2
\omega (t'-t) \; ,
$$
such that $ h(\tau) =  f(\tau+\Theta)\; I(\tau) $.

Using the expansion of the Jacobian elliptic function (see \cite{pbm}) and the
notations (\ref{constant}) we have 
\begin{equation} \label{devf2}
f^2(t)=R^2\left[\frac{E(k)}{K(k)}-k'^2+\frac{2 \pi^2}{k^2
K^2}\sum_{n=1}^\infty \frac{n q^n}{1-q^{2n}}\cos n \Omega t\right] 
\end{equation}
and
\begin{equation} \label{ff1}
ff_1(t+\Theta) =-2\Omega^3\sum_{n=1}^\infty  \frac{n^2
q^n}{1-q^{2n}}\sin n\Omega(t+\Theta) 
\end{equation}
where $ K(k) $ and $ E(k) $ stand for complete elliptic integrals
of first and second kind, respectively.

Then it is straightforward to obtain that:
\begin{eqnarray}\label{Itot}
I(t)&=&\Omega^3 \sum_{n=1}^\infty \frac{n^2 q^n}{1-q^{2n}}\int
d\mu(\omega)\;\left[
{{\cos n\Omega(t+\Theta)- \cos (n \Omega\Theta-2\omega t)
}\over{2\omega+n\Omega}} \right. \cr \cr
&+ & \left. {{\cos(2\omega t+n \Omega \Theta)-\cos n\Omega
(t+\Theta)}\over {2\omega-n\Omega}}\right] 
\end{eqnarray}
We study the behaviour at large $t$ of each term of $I(t)$. For the
first integral we trivially get 
\begin{eqnarray}
\int d\mu(\omega)\; 
{{\cos n\Omega(t+\Theta)}\over {2\omega+n\Omega}}=c_n^{(1)} 
\cos n \Omega(t+\Theta)\cr   \cr
c_n^{(1)}=\int_1^\infty \frac{\sqrt{\omega^2-1}}{2 \pi \omega^2}
\frac{1}{2\omega+n\Omega}\, d\omega \nonumber
\end{eqnarray}
We find for the second integral 
\begin{equation}\label{intasi}
-\int d\mu(\omega) \frac{\cos (n \Omega \Theta-2\omega
t)}{2\omega+n\Omega}\buildrel{t\to \infty}\over =\frac{\sqrt{\pi}}{8}
\frac{\cos(2t-n\Omega
\Theta+\pi/4)}{(2+n\Omega)t^{3/2}}+{\cal O}(\frac{1}{t^{5/2}}) 
\end{equation}
The last integral can be written as
$$
\int d\mu(\omega)\frac{\cos(2\omega t+n \Omega \Theta)-\cos n\Omega
(t+\Theta)}{2\omega-n\Omega}=c_n^{(2)}(t)\; \cos n
\Omega(t+\Theta)-s_n^{(2)}(t) \; \sin n \Omega(t+\Theta) 
$$
Where,
\begin{eqnarray}
c_n^{(2)}(t)&\buildrel{t\to \infty}\over
=&- {\rm v.p.}
\int_{1}^\infty
\frac{\rho(\omega)}{2(\omega-\omega_n)}d\omega+{\cal O}(\frac{1}{t^2})\cr 
s_n^{(2)}(t)&\buildrel{t\to \infty}\over 
=&\frac{\pi}{2}\rho(\omega_n)+{\cal O}(\frac{1}{t^2}) \nonumber
\end{eqnarray}
Here $\rho(\omega)$ and $\omega_n$ are defined as:
$$
\rho(\omega) = \frac{\sqrt{\omega^2-1}}{2 \pi \omega^2} \quad , \quad
\omega_n = \frac{n\Omega}{2} \; .
$$
We now write $I(t)$ as
$$
I(t)=I_{reg}(t)+I_{div}(t) \; .
$$
  $I_{div}(t)$  gives for large $t$ unbounded  contributions
to $\Theta$ while $I_{reg}(t)$ only gives regular terms 
\begin{eqnarray}\label{Idiv}
I_{div}(t)&= &\Omega^3\; \sum_{n=1}^\infty \frac{n^2
q^n}{1-q^{2n}}\left[C_n \cos n\Omega(t+\Theta)-S_n \sin
n\Omega(t+\Theta)\right] \; , \cr \cr
C_n &=& \frac12\ {\rm v.p.}\int_1^\infty
	\left(\frac1{\omega+\omega_n}-\frac1{\omega-\omega_n}\right)
d\mu(\omega) ,\cr \cr
S_n &=& \frac{\pi}{2}\rho(\omega_n)\; .
\end{eqnarray}
We rewrite the renormalization group equations using
$f_1$, $f_2$, $f_3$, $I_{div}$ and $I_{reg}$ as: 
\begin{eqnarray}
\dot\Theta(t) & = & -\left.\frac{\lambda}{W}\right\{ 
I_{div}(t)\, ff_3(t+\Theta)+I_{reg}(t)\, ff_2(t+\Theta) \cr \cr
& -& \frac{d}{dt}\biggl[ g(t)f_3(t+\Theta)\biggr]
-\frac{\Omega'}{\Omega}\, g(t)\, f_1(t+\Theta) \cr \cr
& + & \left.\frac{\Omega'}{\Omega}\, t\biggl[
I_{div}(t)\, ff_1(t+\Theta)-\frac{d}{dt}\biggl(g(t)f_1(t+\Theta)\biggr)\biggr]
\right\} \; , \cr \cr
\dot{R}(t) & = & \frac{\lambda}{W}\Biggl\{ I_{reg}(t)\, ff_1(t+\Theta) 
+ I_{div}(t)\, ff_1(t+\Theta)
-\frac{d}{dt}\biggl[g(t)f_1(t+\Theta)\biggr]\Biggr\}\; .\nonumber
\end{eqnarray}
Since the final resumed solution do not depend on $g(t)$
and since $\eta(t)$ is finite
 we may choose $g(t)$, in order to find a simple expression for $R(t)$,
 as a solution of the equation
$$
I_{div}(t)ff_1(t+\Theta)=\frac{d}{dt}\left[g(t)f_1(t+\Theta)\right]
$$
That is,
$$
g(t)=\frac{1}{f_1(t+\Theta)} \int_{t_0}^t I_{div}(t')ff_1(t'+\Theta)\,
dt' \; .
$$
Now the renormalization group equations become
\begin{eqnarray} \label{rg}
\dot\Theta(t)&=&-\left.\frac{\lambda}{W}\right\{
I_{div}(t)\, ff_3(t+\Theta)+I_{reg}(t)\, ff_2(t+\Theta)  \cr  \cr
& - &\left.
\frac{d}{dt}\biggl[g(t)f_3(t+\Theta)\biggr]-
\frac{\Omega'}{\Omega}\, g(t)f_1(t+\Theta) \right\} \cr   \cr
\dot{R}(t)&=& \frac{\lambda}{W}\, I_{reg}(t)\, ff_1(t+\Theta) \; .
\end{eqnarray}

\section{The order parameter behaviour from the renormalization group
equations}  

We have now the renormalized group equations (\ref{rg}) and we must
remember that the field 
 equations of motion  (\ref{fieldmotion}) are valid
up to ${\cal O}(R_0^3) $.
We have, using the
expansion (\ref{ff1}) and (\ref{Idiv}) in  eq.(\ref{rg})
\begin{eqnarray}\label{idiv}
I_{div}(t)&=&\Omega^3\sum_{n=1}^\infty \frac{n^2 q^n}{1-q^{2n}}\Biggl[C_n
\cos n\Omega(t+\Theta)-S_n \sin n\Omega(t+\Theta)\Biggr] \cr 
I_{reg}(t)&=&I(t)-I_{div}(t) 
\end{eqnarray}
We now introduce the following notations:
\begin{equation} \label{notat}
\alpha_n=\frac 1 q \frac{n^2 q^n}{1-q^{2n}} \quad , \quad
F'(R)=\frac{K(R)^6}{q(R)^2}\frac{R}{(1+R^2)^2}
\end{equation}
With this notations the renormalization group equation (\ref{rg})
for $R(t)$ can be written as: 
\begin{eqnarray}
F'(R)\,\dot R 
=\frac{\pi^6\lambda}2 \sum_{n,m=1}^\infty&&\alpha_n\alpha_m\Biggl\{ 
\pi\rho(\omega_n)\biggl(
\cos2(\omega_m-\omega_n)t-\cos2(\omega_m+\omega_n)t\biggr)
\cr \cr
&&+\mbox{reg}
\int_1^\infty \biggl(
\frac1{\omega-\omega_n}-\frac1{\omega+\omega_n}\biggr)
\biggl(\sin2(\omega_m+\omega)t-\sin2(\omega-\omega_m)t\biggr)
d\mu(\omega)\Biggr\}\nonumber
\end{eqnarray}
Where where by `reg' we  mean the integral regularized by subtracting the
appropriate counterterm contained in $ C_n $ [see eq.(\ref{Idiv})].
We used eqs.(\ref{ff1}-\ref{Itot}) and (\ref{idiv}) and 
we have set $t_0=0$ since the eq.(\ref{fieldmotion}) is invariant under
time translations. Moreover, at this order of $ \lambda $ we can set
$ \Theta = 0 $. Integrating this last expression with respect to $ t $
yields:
\begin{eqnarray} \label{gorda}
F(R(t))-F(R_0)&=&\frac{\lambda\pi^6}4
\sum_{n,m=1}^\infty\alpha_n\alpha_m\Biggl\{
\pi\rho(\omega_n)\biggl(\frac{\sin2(\omega_n-\omega_m)t}{\omega_n-\omega_m}
-\frac{\sin2(\omega_m+\omega_n)t}{\omega_n+\omega_m}\biggr)
\cr
&&+\mbox{reg}\int_1^\infty d\mu(\omega) \biggl(
\frac1{\omega-\omega_n}-\frac1{\omega+\omega_n}\biggr)
\biggl(
\frac{1-\cos2(\omega+\omega_m)t}{\omega+\omega_m}
-\frac{1-\cos2(\omega-\omega_m)t}{\omega-\omega_m}
\biggr)
\Biggr\}
\end{eqnarray}
The terms which look singular for $ n = m $ must be understood in
their limiting value $ n \to m $, which is always regular.

One should notice that, unlike the other terms of the integral, the
last term
gives (for $n$ equal to $m$) a secular term, which is of course canceled by
the last  first  $2\pi\rho(\omega_n) \,t $:
$$
\int_1^\infty d\mu(\omega)\frac{1-\cos2(\omega-\omega_n)t}{(\omega-\omega_n)^2}
-2\pi\rho(\omega_n)t\buildrel{t\to \infty}\over={\rm v.p.}\int_1^\infty 
    \frac{d\mu(\omega)}{(\omega-\omega_n)^2}
    +{\cal O}(\frac1{t^2})
$$
The equations of motion (\ref{fieldmotion})-(\ref{nucleo}) are
valid to order $ R_0^3 $ and $ q = O(k^2) = O(R_0^2) $ for small $
R_0 $. Therefore, we can keep  the terms
in  the sum (\ref{gorda}) where $n$ and $m$ are equal to $1$. 
We find from eq.(\ref{gorda}) after non-trivial cancelations,

\begin{eqnarray} \label{final}
F(R(t))-F(R_0)&=&\frac{\lambda\pi^6}4
\Biggl\{
\pi\rho(\omega_1)\biggl(2t
-\frac{\sin4\omega_1t}{2\omega_1}\biggr)
\cr \cr
&&+\mbox{reg}\int_1^\infty d\mu(\omega) \biggl(
\frac1{\omega-\omega_1}-\frac1{\omega+\omega_1}\biggr)
\biggl(
\frac{1-\cos2(\omega+\omega_1)t}{\omega+\omega_1}
-\frac{1-\cos2(\omega-\omega_1)t}{\omega-\omega_1}
\biggr)\; ,
\Biggr\}
\end{eqnarray}
where $ \omega_1 = \Omega/2 = {{\pi}\over { 2 K(k) }}\sqrt{1+R^2} $. 

In order to find the field amplitude for $ t \to \infty $, we average
over an oscillation period 
\begin{eqnarray}
F(R_\infty)-F(R_0)=-\frac{\pi^5\lambda}8\ {\rm v.p.}\int_1^\infty d\omega\,
\frac{\sqrt{\omega^2-1}}{\omega^2}
\left\lbrack\frac{1}{(\omega+\omega_1)^2}+
\frac{1}{(\omega-\omega_1)^2}\right\rbrack \nonumber
\end{eqnarray}
The last integral gives:
\begin{eqnarray}\label{Rinfi}
F(R_\infty)-F(R_0)=-\frac{\pi^5\lambda}{\omega_1^2}\left\lbrack
\frac{\omega_1^2+2}{\omega_1\sqrt{\omega_1^2-1}}
{\rm Arg}\cosh\omega_1-2\right\rbrack \; ,
\end{eqnarray}

One can easily see that the expression between the brackets is
positive. Since $ F'(R) $ is positive $ R_\infty $ is smaller than
$R_0$. This result was expected since we have a dissipative
behaviour. Namely, part of the energy initially in the zero mode is
spent in particle production.

More explicitly,
\begin{equation} \label{funF}
F(R) \equiv \int_{\infty}^R \frac{K(k(R))^6}{q(k(R))^2}\frac{R\,
dR}{(1+R^2)^2}= \int_{1/\sqrt2}^{k(R)} k \; dk \; {{K(k)^6}\over
{q(k)^2}} \; .
\end{equation}
where $ k(R) $ is defined by eq.(\ref{kdeR}).

For late $ t $, $ R(t) - R_{\infty} $ is small and we can write
$$
F(R(t))-F(R_\infty)=F'(R_\infty)(R(t)-R_\infty) \; .
$$
We can then evaluate $ F(R(t)) $ for late $ t $ from eq.(\ref{final}) using
eq.(\ref{intasi}), 
$$
F(R(t))-F(R_\infty)
=\pi^6\; \lambda \frac {\sqrt\pi}4
\int_t^\infty\frac{\cos(2t'+\pi/4)\sin\Omega t'}{(2+\Omega)t'^{3/2}}\,
dt' 
$$
 We find equating both results  the behaviour of $R(t)$ for late $t$:
\begin{eqnarray}
R(t)-R_\infty
=\frac{\lambda\sqrt\pi}{32}\frac{R_\infty^3}{1-3R_\infty^2}
\frac{\Omega\cos\Omega t \; \cos(2t+\pi/4)+\sin\Omega t
\; \sin(2t+\pi/4)}{(\Omega+2)^2(\Omega-2)\; t^{3/2}} +O(t^{-2})
\end{eqnarray}

Eqs.(\ref{Rinfi}) and  (\ref{funF}) define $ R_\infty $ as a
function of $ R_0 $  and $ \lambda $.

In the present case it is very convenient to express the elliptic
functions in powers of the elliptic nome $ q = e^{-\pi K'/K} $ since $
q $ is always small in our context. Namely,
$ 0 \leq q \leq e^{-\pi}  = 0.0432139\ldots $ for $ 0 \leq R_0 \leq
\infty $. We have for example,
\begin{eqnarray}
K(k) &=& \frac{\pi}2 \left[ 1 + 4 q + 4q^2 +{\cal O}(q^4)\right] \cr \cr
k^2 &=& 16q  \left[ 1 - 8 q + 44 q^2 + {\cal O}(q^3)\right] \; .\nonumber
\end{eqnarray}
The function $ F(R) $ becomes 
\begin{equation}\label{Fexpli}
F(R) = - {{\pi^6}\over 8}\; \left[ \frac1{q} - 8 \log q -12 q + {\cal O}(q^2)
\right] + {\cal C}\; .
\end{equation}
where the constant $  {\cal C} $ is given by
$  {\cal C} = {{\pi^6}\over 8}\; e^{\pi} \; \left[1 + 8 \pi e^{-\pi} -
12   e^{-2\pi} + {\cal O}(  e^{-3\pi} ) \right] = 5738.9\ldots $

Eq.(\ref{Rinfi}) and (\ref{Fexpli})
indicates us that $q_{\infty}\equiv q(R_{\infty})$ and $q_0 \equiv
q(R_0)$ are related as follows,
$$
q_{\infty} - q_0 = -c \; \lambda + {\cal O}( \lambda^2) \; .
$$
where,
$$
c =  \frac{8\,q_0^2}{\pi\omega_1^2}\left\lbrack
\frac{\omega_1^2+2}{\omega_1\sqrt{\omega_1^2-1}}
{\rm Arg}\cosh\omega_1-2\right\rbrack [1 - 8 q_0 + {\cal O}( q_0^2) ]\; .
$$
More explicitly, for $R_0\to 0$ we find:
$$
c\simeq\frac 8\pi q_0^2 \simeq\frac{R_0^4}{128\pi}+{\cal O}(R_0^6)
$$
That is:
$$
R_\infty\simeq R_0\biggl(1-\lambda\frac{R_0^2}{8\pi}\biggr) +
{\cal O}(\lambda R_0^4) \; .
$$

\acknowledgements

We would like to thank D. Boyanovsky,
B. Delamotte and I. Egusquiza for useful discussions.

\end{document}